\documentclass[preprint,NumberedRefs]{JASA}

\usepackage{subcaption}
\usepackage{array}

\begin{document}

\title[preprint submitted to JASA]{Ultrasonic characterization of generally anisotropic elasticity implementing optimal zeroth-order elastic bounds and a wave-fitting approach}
\author{Diego A. Cowes}
\email{diegocowes@cnea.gob.ar}
\affiliation{Departamento ICES, Comisión Nacional de Energía Atómica, Villa Maipú, B1650, Argentina}
\affiliation{Instituto Sabato, Universidad Nacional de San Martín, Villa Maipú, B1650, Argentina}

\author{Juan I. Mieza}
\affiliation{Instituto Sabato, Universidad Nacional de San Martín, Villa Maipú, B1650, Argentina}
\affiliation{Hidrógeno en materiales, Comisión Nacional de Energía Atómica, Villa Maipú, B1650, Argentina}

\author{Martín P. Gómez}
\affiliation{Departamento ICES, Comisión Nacional de Energía Atómica, Villa Maipú, B1650, Argentina}
\affiliation{Instituto Sabato, Universidad Nacional de San Martín, Villa Maipú, B1650, Argentina}
\affiliation{Departamento de Mecánica, Universidad Tecnológica Nacional FRD, Campana, B2804, Argentina}

\preprint{Diego A. Cowes, JASA}	

\date{\today}

\begin{abstract}
The elastic behavior of materials is of critical importance for the design, fabrication, and testing of industrial and structural components. The ease with which the wave angle of incidence can be varied makes ultrasonic techniques well suited for the characterization of anisotropic materials, whose properties are direction-dependent. This work aims to develop an ultrasonic goniometry method in which a wave is transmitted through a sample while scanning over spherical coordinates. A plane-wave model is formulated that accounts for fluid–solid interfaces and is applicable to a wide range of sample thicknesses. The model assumes general anisotropy, enabling the characterization of materials with symmetries up to triclinic, and does not require precise sample alignment. Specially designed transducers support the plane-wave approximation, thereby avoiding the need for more computationally expensive finite-beam models. Furthermore, implementation of the forward model on GPU architectures significantly reduces the computational cost associated with the numerous evaluations required during the waveform fitting inversion. The introduction of optimal zeroth-order bounds is used to tightly delimit the search space, and an isotropic self-consistent solution is shown to provide an effective initial guess. Finally, measurements on plate-like samples are compared with the literature and diffraction-based methods.   
\end{abstract}


\maketitle


\section{\label{sec:1} Introduction}
The determination of elastic constants is commonly performed using ultrasonic methods due to their flexibility and their intrinsic ability to characterize anisotropic materials. Among these techniques, resonance-based approaches, such as Resonant Ultrasound Spectroscopy (RUS), are widely used, but they require samples with specific geometries, often demanding extensive preparation \citep{Migliori2005}.

Traveling-wave ultrasonic methods provide an alternative by transmitting pulses through a sample along multiple directions. Early implementations relied on contact transducers, which also necessitated specialized sample preparation, such as machining flat faces normal to each propagation direction \citep{Wachtman1960, Kriz1979, Papadakis1991}. The emergence of fiber-reinforced and other advanced anisotropic materials increased the need to characterize thin plates along many directions. This demand fostered the development of immersion-based traveling-wave methods \citep{Markham1970, Hosten1987, Rokhlin_1989} , where fluid coupling enables continuous variation of the incidence angle. As a result, only a pair of parallel faces is typically required, simplifying sample preparation and allowing ultrasonic inspection of plate materials during fabrication or in service.

When the material macroscopic symmetry is known, elastic constants can be directly inferred from phase velocities measured at selected directions \citep{truell1969ultrasonic}. However, phase-velocity parametrization in immersion goniometry is non-trivial and error-prone, especially when the sample thickness approaches the ultrasonic wavelength \citep{lavrentyev1997phase, Wang2003}, voiding bulk wave assumption. An alternative strategy is to fit a plane-wave propagation model directly to the measured waveforms \citep{Leymarie2002}, spectra \citep{Castaings2000} or scattering coefficients \citep{Lobkis1996} using nonlinear optimization, often implemented via metaheuristic iterative methods that require well-defined initial guesses and parameter bounds.

To reduce the inversion complexity, measurements are sometimes restricted to symmetry planes\citep{Rokhlin1992, Aristgui1997, Castaigns_Materials}, where the mechanical response depends only on a subset of elastic constants \citep{nayfeh1995wave, Chimenti2011-xj}. This strategy, however, requires prior knowledge of the material macroscopic symmetry and precise alignment of the specimen during measurement. More recent work has sought to extend the applicability of these methods to new materials \citep{Masud2024, Poudrel2024, LeBourdais2024}; however, it does not address several of the shortcomings highlighted above. 

The objective of this work is to develop a general framework for determining the elastic constants of anisotropic crystalline materials using ultrasonic goniometry through waveform fitting. The use of specially designed transducers permits the plane wave assumption, instead of a more computationally expensive finite beam model \citep{Cowes2024}. The plane wave model, which takes into account both fluid-solid interfaces, permits fitting for both thick and thin specimens, where the bulk wave assumption would fail. Moreover, no macroscopic symmetry assumptions are enforced so that the method is applicable for up to general anisotropy (triclinic), and precise sample alignment becomes unnecessary. While considering general anisotropy is more computationally expensive than restricting the symmetry, this is offset by the usage of GPUs which takes advantages of parallelizable operations. 

A key novelty lies in introducing optimal zeroth-order elastic bounds for delimitating the search space, and the usage of a self-consistent initial guess corresponding to the isotropic state, using the framework developed by Lobos et al. \citep{Lobos2016,lobos2018homogenization} extending the Hashin-Shtrikman theory \citep{Hashin1962}. This combination enables robust elastic characterization of plate-like materials, independent of sample thickness, and without requiring prior knowledge of the macroscopic symmetry or symmetry directions. Measurements were carried for reference single crystals of known properties as well as polycrystalline metallic plates of different thicknesses. The measurements were compared with the elastic constants computed from neutron and X-ray diffraction with homogenization methods. 

The paper is organized as follows: Section \ref{sec:2} describes the theoretical basis which include the plane wave model in subsection \ref{subsec:2:1}, and the elastic bounds in subsection \ref{subsec:2:2}. Section \ref{sec:3} describes the experimental method, including details about the ultrasonic goniometry, the derivative-free optimization, and the sample selection. Section \ref{sec:4} shows the results that are discussed in Section \ref{sec:5}, followed by some concluding remarks presented on Section \ref{sec:6}.

\section{Theoretical Basis}\label{sec:2}
\subsection{Ultrasonic plane wave model}\label{subsec:2:1}
In this work, the transmission coefficient of an immersed plate is calculated according to the work of Nayfeh and Chimenti \citep{Nayfeh1989, nayfeh1995wave}, developed for the non destructive testing of fiber reinforced materials but reformulated here for generally anisotropic homogeneous materials. 

\subsubsection{Wave equations and coordinate rotation}
The dynamic behavior of a linear elastic generally anisotropic solid can be expressed by the equation of motion
\begin{equation}\label{eq:w1}
    \frac{\partial \sigma_{ij}'}{\partial x_j'}=\rho \frac{\partial^2u_i'}{\partial t^2},
\end{equation}
the general stress-strain constitutive relations
\begin{equation}\label{eq:w2}
    \sigma_{ij}'= c_{ijkl}'e_{kl}',
\end{equation}
and the linearized strain-displacement relations
\begin{equation}\label{eq:w3}
    e'_{kl}=\frac{1}{2}(\frac{\partial u'_l}{\partial x'_k}+\frac{\partial u'_k}{\partial x'_l}),
\end{equation}
where $\sigma_{ij}'$, $u_i'$, $c_{ijkl}'$, $e_{kl}'$, and $\rho$ are the stress tensor, displacement vector, stiffness tensor, strain tensor and density, respectively.

\begin{figure}
    \centering
    \includegraphics[width=0.8\linewidth]{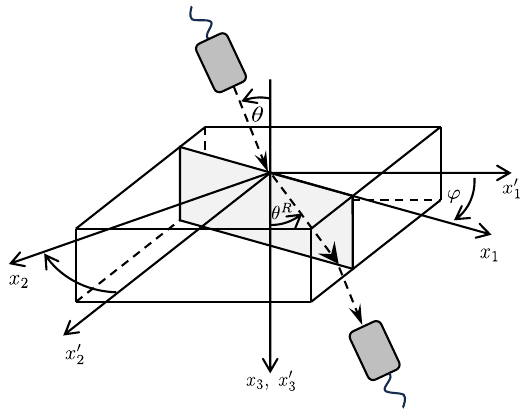}
    \caption{Diagram of the ultrasonic goniometry experiment. The azimuthal rotation by $\varphi$ determines the plane of incidence. The polar rotation by $\theta$ determines the angle of incidence, while $\theta^R$ is the angle of refraction.}
    \label{Figure1}
\end{figure}

As depicted in Fig. \ref{Figure1}, the primed coordinate system $(x_1^{'}, x_2^{'}, x_3^{'})$ corresponds to the one aligned with the material directions. The unprimed coordinate system corresponds to an azimuthal rotation $\varphi$ around the plate normal $x_3^{'}$, which is chosen so that the incident and refracted wave vectors are included in the $(x_1, x_3)$ plane. This transformation can be performed with a tensor product or as the Bond method \citep{Bond1943} via matrix multiplication, 
\begin{equation}\label{eq:w5}
    C=M(\varphi)C'M(\varphi)^T
\end{equation}
where $M(\varphi)$ is the Bond rotation matrix and $C$ is the stiffness tensor represented in Voigt notation which, for a generally anisotropic solid, is composed of 21 independent elastic constants. Eq. \ref{eq:w2} can be rewritten as
\begin{equation}\label{eq:w6}
    \begin{bmatrix}       
    \sigma_{11} \\\sigma_{22}\\\sigma_{33} \\\sigma_{23} \\\sigma_{13}\\ \sigma_{12}
    \end{bmatrix}=
    \begin{bmatrix}
    C_{11}&C_{12}&C_{13}&C_{14}&C_{15}&C_{16}\\
    \,    &C_{22}&C_{23}&C_{24}&C_{25}&C_{26}\\
    \,    &\,    &C_{33}&C_{34}&C_{35}&C_{36}\\
    \,    &\,     &\,    &C_{44}&C_{45}&C_{46}\\
   \,     &\text{sym}  &\,    &\,   &C_{55}&C_{56}\\
    \,    &\,    &\,    &\,    &\,    &C_{66}
    \end{bmatrix}
    \begin{bmatrix}
        e_{11}\\e_{22}\\e_{33}\\\gamma_{23}\\\gamma_{13}\\\gamma_{12}
    \end{bmatrix}
\end{equation}

\subsubsection{Fluid immersed plate }
\begin{figure}
    \centering
    \includegraphics[width=0.8\linewidth]{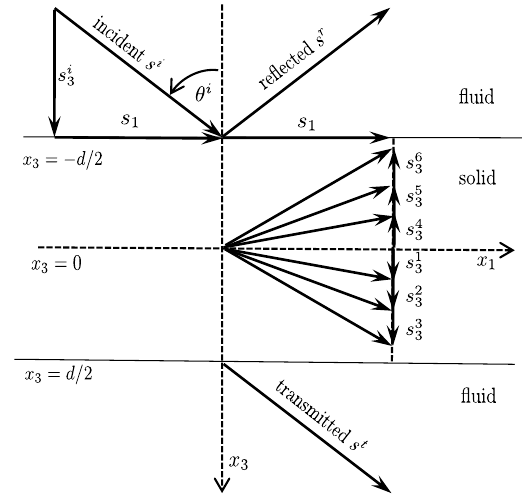}
    \caption{Partial waves in a fluid immersed triclinic plate.}
    \label{Figure2}
\end{figure}

When a triclinic plate is immersed in a fluid, as depicted in Fig. \ref{Figure2}, an incident wave gives rise to a reflected wave a transmitted wave and six refracted partial waves, three traveling in the positive $x_3$ direction and three traveling in the negative $x_3$ direction. For each direction there exists a quasi-longitudinal, a quasi-shear horizontal and a quasi-shear vertical wave.

The Christoffel equation can be solved to find phase velocities $v$ for a propagation direction $n_j$ in homogeneous linearly elastic  generally anisotropic media. However,  at a fluid-solid interface, due to refraction, it's not possible to know $n_j$ a priori. Conversely, if the wave impinges from the fluid at an angle $\theta$ from the interface normal, according to Snell's law the wavevector component $k_1$, parallel to the interface, is common to both media, and the unknowns become the vertical wavevector components $k_3^q$ or equivalently, the vertical slowness components $s_3^q$, where $k_j=\omega s_j$, $\omega$ is the angular frequency, and $|s_j|=1/v$.

By virtue of the coordinate transformation in Eq. \ref{eq:w5}, motion in the plane $(x_1,x_3)$ is independent of $x_2$ and the solutions for the displacement equations can be sought in the form 

\begin{equation}\label{eq:w7}
    u_j=U_j e^{i\omega(s_1 x_1+s_3 x_3-t)}, \,\,\,\,  j=1,2,3.
\end{equation}

 Substituting Eq. \ref{eq:w7} into Eq. \ref{eq:w1} leads to three coupled equations 
\begin{equation}\label{eq:w8}
K_{ij}(s_3)U_j=0, \,\,\,\,  i,j=1,2,3.
\end{equation}
This is the alternative form of the Christoffel equation, which relates the slowness horizontal component (known) with the vertical components (unknown). For the existence of nontrivial solutions, the determinant of $K_{ij}$ must vanish, giving a sixth-degree polynomial. For each $s_3^q$, Eq. \ref{eq:w8} is solved to determine the normalized displacement amplitudes $\mathcal{U}_2=U_2/U_1$ and $\mathcal{U}_3=U_3/U_1$, which define wave polarization. Subsequently, normalized stress amplitudes $\mathcal{T}_{ij}$ are derived from Eqs. \ref{eq:w2} and \ref{eq:w3}. Dropping the common factor $e^{i\omega(s_1 x_1-t)}$ and combining normalized displacements and stresses with Eq. \ref{eq:w7}, the formal solutions for the displacements become
\begin{equation}\label{eq:w10}
(u_1,u_2,u_3)=\sum_{q=1}^6(1,\mathcal{U}_2^q,\mathcal{U}_3^q)U_1^q  e^{i\omega s_3^q x_3}, 
\end{equation}
and for the stresses
\begin{equation}\label{eq:w11}
(\sigma_{33},\sigma_{13},\sigma_{23})=\sum_{q=1}^6i\omega(\mathcal{T}_{33}^q,\mathcal{T}_{13}^q,\mathcal{T}_{23}^q)U_1^q e^{i\omega s_3^q x_3}.
\end{equation}

Fluid media do not resist shear deformation, so that the complete formal solutions for fluids result in
\begin{equation}\label{eq:w12}
(u_1,u_3,\sigma_{33})=\sum_{q=1}^2(1,\bar{\mathcal{U}}_3^{q},i\omega\bar{\mathcal{T}}_{33})\bar{U}_1^q e^{i\omega \bar{s}_3^{q} x_3}
\end{equation}
where $\bar{s}_3^{q}=(-1)^{q+1}\sqrt{\frac{1}{v_f^2}-s_1^2}$, $\bar{\mathcal{U}}^q=\frac{\bar{s}_3^{q}}{s_1}$, $\bar{\mathcal{T}}_{33}=\frac{\rho_{f}}{s_1}$ and $v_f$ is the sound velocity in the fluid.

Fluid-solid interfaces are characterized by the continuity conditions
\begin{equation}\label{eq:w13}
    \bar{u}_3=u_3,\quad \bar{\sigma}_{33}=\sigma_{33}, \quad \sigma_{13}=\sigma_{33}=0,
\end{equation}
where a bar indicates a quantity in the fluid, and no bar indicates a quantity in the solid. These conditions imposed at the solutions and specialized at $x_3=-d/2,d/2$, lead to a system of equations
\def\vd{5pt}
\begin{equation}\label{eq:w14}
\begin{adjustbox}{width=0.48\textwidth}
$\begin{bmatrix}
    \mathcal{U}_3^1E^1 &\mathcal{U}_3^2E^2 &\mathcal{U}_3^3E^3 &\mathcal{U}_3^4E^4 &\mathcal{U}_3^5E^5 &\mathcal{U}_3^6E^6 &-\bar{\mathcal{U}}_3^2&0\\[\vd]
    \mathcal{T}_{33}^1E^1 &\mathcal{T}_{33}^2E^2 &\mathcal{T}_{33}^3E^3 &\mathcal{T}_{33}^4E4 &\mathcal{T}_{33}^5E^5 &\mathcal{T}_{33}^6E^6 &-\bar{\mathcal{T}}_{33} &0\\[\vd]
    \mathcal{T}_{13}^1E^1 &\mathcal{T}_{13}^2E^2 &\mathcal{T}_{13}^3E^3 &\mathcal{T}_{13}^4E4 &\mathcal{T}_{13}^5E^5 &\mathcal{T}_{13}^6E^6 &0 &0\\[\vd]
    \mathcal{T}_{23}^1E^1 &\mathcal{T}_{23}^2E^3 &\mathcal{T}_{23}^3E^3 &\mathcal{T}_{23}^4E4 &\mathcal{T}_{23}^5E^5 &\mathcal{T}_{23}^6E^6 &0 &0\\[\vd]
    \mathcal{U}_3^1\hat{E}^1 &\mathcal{U}_3^2\hat{E}^2 &\mathcal{U}_3^3\hat{E}^3 &\mathcal{U}_3^4\hat{E}^4 &\mathcal{U}_3^5\hat{E}^5 &\mathcal{U}_3^6\hat{E}^6 &0 &-\bar{\mathcal{U}}_3^1\\[\vd]
    \mathcal{T}_{33}^1\hat{E}^1 &\mathcal{T}_{33}^2\hat{E}^2 &\mathcal{T}_{33}^3\hat{E}^3 &\mathcal{T}_{33}^4\hat{E}^4 &\mathcal{T}_{33}^5\hat{E}^5 &\mathcal{T}_{33}^6\hat{E}^6 &0 &-\bar{\mathcal{T}}_{33}\\[\vd]
    \mathcal{T}_{13}^1\hat{E}^1 &\mathcal{T}_{13}^2\hat{E}^2 &\mathcal{T}_{13}^3\hat{E}^3 &\mathcal{T}_{13}^4\hat{E}^4 &\mathcal{T}_{13}^5\hat{E}^5 &\mathcal{T}_{13}^6\hat{E}^6 &0 &0\\[\vd]
    \mathcal{T}_{23}^1\hat{E}^1 &\mathcal{T}_{23}^2\hat{E}^2 &\mathcal{T}_{23}^3\hat{E}^3 &\mathcal{T}_{23}^4\hat{E}^4 &\mathcal{T}_{23}^5\hat{E}^5 &\mathcal{T}_{23}^6\hat{E}^6 &0 &0
\end{bmatrix}    
\begin{bmatrix}
    U_1^1\\[\vd] U_1^2\\[\vd] U_1^3\\[\vd] U_1^4\\[\vd] U_1^5\\[\vd] U_1^6\\[\vd]\bar{U}_1^R\\[\vd] \bar{U}_1^T
\end{bmatrix}
=
\begin{bmatrix}
    \bar{\mathcal{U}}_3^1 \bar{U}_1^I\\[\vd] \bar{\mathcal{T}}_{33}\bar{U}_1^I \\[\vd]0 \\[\vd]0 \\[\vd]0 \\[\vd]0 \\[\vd]0 \\[\vd]0
\end{bmatrix}$
\end{adjustbox}
\end{equation}
where $\hat{E}^q=(E^q)^{-1}=e^{i\omega s_3^q d/2}$.

This system can be solved for the reflection coefficient $R=\bar{U}_1^R/\bar{U}_1^I$ or the transmission coefficient $T=\bar{U}_1^T/\bar{U}_1^I$.

\subsubsection{Transmitted field}
The transmitted field is defined as the convolution in the frequency domain of the transmission coefficient and the system spectrum, $P$, as
\begin{equation}
T_f(\varphi,\theta,  \omega)=T(\varphi,\theta,  \omega)P(\omega)e^{i\phi},
\end{equation}
where $e^{i\phi}$ is a phase correction term accounting for the variation in travel distance when introducing the plate. The ultrasonic inversion can be performed by fitting the transmission coefficient $|T|$, although the transmitted field $|T_f|$ is preferred because the pulse deconvolution can lead to numerical errors \citep{Castaings2000}.

Alternatively, the temporal signal can be simulated by performing an inverse Fourier transform as
\begin{equation}\label{eq:w15}
s(\varphi,\theta, t)=\int_{-\infty}^{\infty} T_f(\varphi,\theta,  \omega)e^{-i\omega td\omega}.
\end{equation}
In this work, the latter was used for the inversion because it naturally includes phase information.

\begin{figure}
    \centering
    \includegraphics[width=0.95\linewidth]{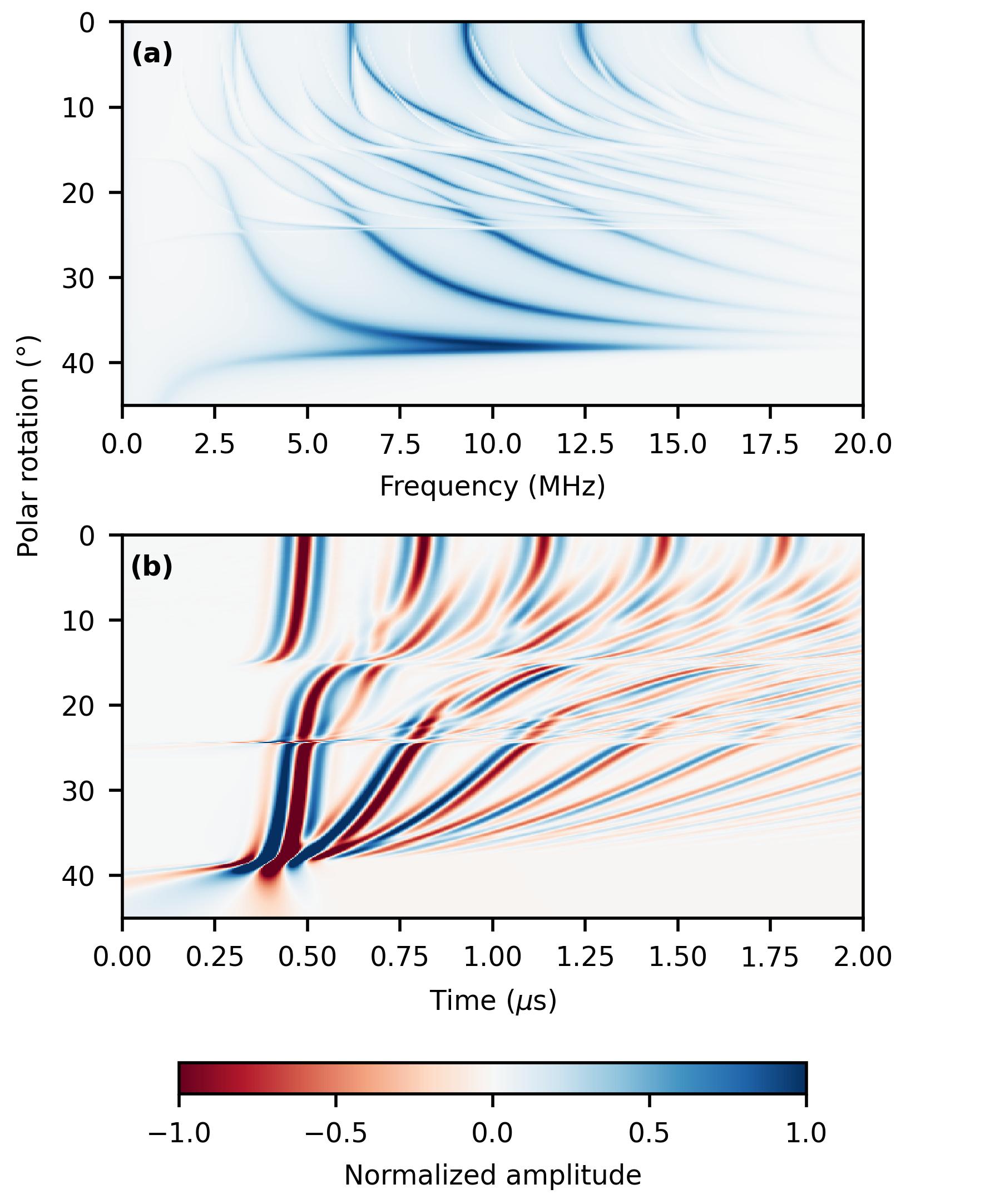}
    \caption{Plane wave simulation of a polar scan on a triclinic steel plate. (a) Magnitude of the transmission coefficient. (b) Transmitted signal. (Color online).}
    \label{Figure3}
\end{figure}

A simulation for a triclinic plate with a thickness of 1 mm is shown in Fig. \ref{Figure3}. The transmission coefficient $|T|$ and the transmitted signal $s$ are shown for a single polar scan at fixed azimuthal angle.

\subsection{Elastic Bounds}\label{subsec:2:2}
Effective elastic behavior of polycrystalline materials is energetically bounded by the strain energy density $W$ as, 
\begin{equation}\label{eq:b1}
    W^- \leq \bar{W} \leq W^+,\  \text{where}\ \  W=\frac{1}{2}e_{ij}c_{ijkl}e_{kl}.
\end{equation}
This definitions leads to the existence of an upper and a lower stiffness tensor as,
\begin{equation}
    \bar{e}_{ij}c^-_{ijkl}\bar{e}_{kl} \leq \bar{e}_{ij}\bar{c}_{ijkl}\bar{e}_{kl} \leq \bar{e}_{ij}c^+_{ijkl}\bar{e}_{kl},
\end{equation}
where the $+$ and $-$ signs denote the upper and lower bounds respectively, and $\bar{e}_{ij}$ represents the effective strain tensor.

Zeroth-order bounds are independent of microstructural statistical information, such as crystallographic texture or volumetric fractions, and therefore consist of isotropic tensors. Moreover, these bounds energetically encompass all possible realizations (microstructures) of the polycrystalline material under investigation.  
The upper bound is computed by finding the tensor with minimum trace complying with Eq. \ref{eq:b1} \citep{Nadeau2001}, and the lower bound is computed with the same procedure but employing the compliance tensor \citep{Lobos2016}. This operation needs the single crystal stiffness tensor as an input, which can usually be found in the literature. In this work, only single phase materials were tested. If multi phase-materials are considered the zeroth-order bounds can also be computed by taking into account the volume fraction for each phase.

In practice, each component of the effective stiffness tensor is enclosed by the relations defined by Lobos et al. \citep{Lobos2016}. For the diagonal components in Voigt notation as,
\begin{equation}\label{eq:b3}
    C^-_{\alpha\alpha} \leq \bar{C}_{\alpha\beta} \leq C^+_{\alpha\alpha} \quad \alpha= \beta,
\end{equation}
and for the off-diagonal components as,
\begin{equation}\label{eq:b4}
    \Gamma^-_{\alpha \beta} \leq \bar{C}_{\alpha \beta} \leq \Gamma^+_{\alpha \beta} \quad \alpha \neq \beta
\end{equation}
\begin{equation}
    \Gamma^\pm_{\alpha \beta}=\frac{1}{2}\Bigl( C^-_{\alpha\beta}+C^+_{\alpha\beta}\pm \sqrt{(C^+_{\alpha\alpha}-C^-_{\alpha\alpha})(C^+_{\beta\beta}-C^-_{\beta\beta})}\ \Bigr)
\end{equation}

If microstructural statistical information is known, higher order (and tighter) bounds may be estimated, such as Voigt, Russ and Hashin-Shtrikman. If the crystallographic texture is assumed to be random, an isotropic self consistent solution ($C_0^{sc}$) can be computed \citep{LobosFernndez2018}. This approximation can be used as a sensible initial guess during the inversion procedure for a material with unknown microstructure. 

\section{Experimental method}\label{sec:3}
\subsection{Ultrasonic goniometry}
Stiffness and, consequently, ultrasonic fields are direction dependent quantities which shall be sampled accordingly. An ultrasonic goniometer has been developed for the present work which allows the polar and azimuthal rotation of a fluid immersed sample located between a pair of ultrasonic transducers, as depicted in Fig. \ref{Figure1}. Data acquisition consists of several polar scans taken at different azimuthal angles which constitutes a spherical sampling grid. The step of the grid is critical in limiting experimental and inversion time, but must be small enough to correctly sample the orientation dependent field. 
A strategy proposed by Lan\citep{Lan2018} consists on treating the phase velocity as a spherical surface which can be decomposed in spherical harmonics. This shows that the velocity is a band limited signal which does not have energy above the 6th degree. From this, the author concludes that 12 azimuthal planes, at 3 polar angles, suffice for the correct acquisition of this parameter.
Nevertheless, the present work does not directly measure the phase velocity, but the transmitted ultrasonic  field. This means that the traveling pulse is subjected to refraction and reflection at the fluid-solid interfaces, which results in a more complex signal which requires a finer sampling step in the polar direction. Because of this, a grid of 200 polar angles at 12 azimuthal angles was selected.

The goniometry apparatus developed for this work was automated so that it is able to measure a single sample in less than two minutes.

\subsection{Bandwidth of validity}

Plane wave assumption is required for the application of the ultrasonic model presented in subsection \ref{subsec:2:1} . Ultrasonic goniometry experiments usually generate a lateral field displacement due to leaky guided wave propagation, which finite transducers fail to completely acquire, voiding plane wave validity. A finite beam model can simulate such a setup but commonly requires the double integration for all the wavevectors of interest and as such it is computationally prohibitive for the numerous evaluations during the inversion.  Dedicated PVDF transducers have been developed for the use in ultrasonic goniometry experiments by an optimization process which found the transducer's shape and position which better resemble plane wave propagation. The fabricated transducers were successfully validated by experiments in anisotropic media \citep{Cowes2024}. 

Plane wave assumption may also be defied by transducer diffraction which becomes predominant at low frequencies. Conversely, at high frequencies grain size becomes comparable to wavelength making grain scattering control wave propagation, which challenges the media homogeneity assumption. More complex models can account for grain shape and size while simultaneously considering elasticity \citep{Sha2018}, but become computationally expensive as well. 

The Gaussian frequency response of the damped piezoelectric transducer naturally weighs the frequencies of interest diminishing the influence of grain scattering at high frequencies and beam diffraction at low frequencies. Care should be taken while selecting central frequency. In this work a central frequency of 10 MHz was chosen and high pass and low pass filters (at 10 and 25 MHz, respectively) were employed to further limit the experiment bandwidth.

\subsection{Waveform fitting inversion}
\begin{figure}
    \centering
    \includegraphics[width=0.9\linewidth]{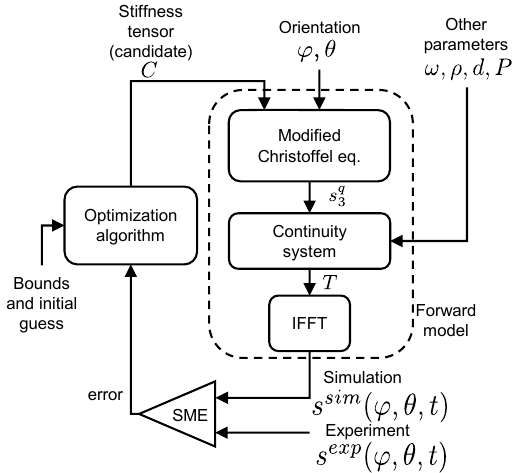}
    \caption{Block diagram for the inversion process.}
    \label{Figure4}
\end{figure}

Because the equations that relate elastic constants to transmitted ultrasonic fields are not analytically invertible, a fitting approach is needed by which the inputs of a forward model are iterated until the error between the experimental data and the model is minimized.
When the problem is non-linear, non-differentiable or non-continuous, a direct search approach is required by which a metaheuristic strategy employs a set of rules that try to lead to the global minimum. While these methods do not guarantee convergence, they do not require information of the objective function other than its inputs and outputs. Some of these algorithms are based on population strategies by which the cost of each individual impacts the choice of new individuals. This, coupled with some degree of stochasticity, results in a better exploration of the search space and a better probability of finding the global minimum. Among these, Pattern Search is an algorithm which makes small exploratory moves searching for patterns which guide the consequent movement. It has been proven to be effective for multi-variable global optimization, even for objective functions with sharp transitions \citep{Hooke1961}. The Python Pymoo library \citep{pymoo} implementation was used for the optimization.
The objective function is defined for a candidate matrix ($C_{ij}$) as the squared mean error (SME)
\begin{equation}
    F(C_{ij})=\frac{1}{N}\sum_{i=0}^N(s_i^{sim}(C_{ij})-s_i^{exp})^2,
\end{equation}
where, $s^{sim}$ and $s^{exp}$ are the simulated and experimental signals respectively, the subindex $i$ goes over the variables $\varphi,\theta,t$, and $N$ is the total sampled points. A small proportional complex term is added to each elastic constant to account for attenuation (scattering, absorption, and others). Adding this term improves fitting, but the response is relatively insensitive to its magnitude \citep{Hosten2008}. The inversion process is represented as a block diagram in Fig. \ref{Figure4}.

\subsection{Computation}
The forward model consists of the computation of a single individual which starts from a set of elastic constants and ends with a set of transmitted signals, by following Eqs. \ref{eq:w5} --\ref{eq:w15}. For a single individual, the calculations must be conducted for each combination of  $\varphi$, $\theta$ and $\omega$. 

Since the code consists of a great number of repeated operations, it is highly parallelizable which means a that it can be carried out in commercial graphical processing units (GPUs) with great time savings. Most of the calculation of the forward model was implemented for GPU architecture with the use of the Python CuPy library \citep{cupy_learningsys2017}, which is a drop-in replacement of the NumPy library \citep{2020NumPy-Array}. The optimization of the forward model is crucial to limit inversion time, because the optimization algorithm requires the computation of approximately 5000 individuals to reach a global minimum. GPU implementation provided a 10-fold reduction in computation time. 

Zeroth-order bounds and the isotropic self consistent solution (initial guess) were computed for each material following the work by Lobos et. al in \citep{Lobos2016} and \citep{lobos2018homogenization}, respectively, but adapted to be run on Python. This information was used by the optimization algorithm to guide the inversion procedure.

\subsection{Sample selection}
For this investigation, four samples were selected: two single-crystal silicon (Si) wafers and two Zircaloy-4 (Zry) plates. Silicon was chosen as a reference material because its elastic constants are well established in the literature from direct measurement methods. One sample was selected by its low thickness ($\approx$ 0.3 mm) to test the methods effectiveness when echo superposition is strong, were the bulk wave assumption would fail. The other sample was selected for its crystallographic orientation, having the ${<}111{>}$ direction normal to the surface. This orientation lacks in-plane mirror symmetry, challenging an assumption commonly made in the literature. The two Zry plates were selected as polycrystalline samples to evaluate both the effectiveness of the defined bounds and the ability of the method to operate across different sample thicknesses (approximately 1 mm and 5 mm, respectively). 

Table \ref{Table1} describes the samples by name, material, thickness, density, and computed bounds. The thickness was determined with a digital micrometer and the density was measured by gas pycnometry. All samples have a surface of at least 20 $\mathrm{cm}^2$. Zeroth-order bounds, being isotropic are defined by two independent variables, in this case $C_{11}$ and $C_{44}$.

\begin{table}[h!]
\caption{Sample details.}
\label{Table1}
\begin{adjustbox}{width=\columnwidth}
\begin{tabular}{@{}cccccccc@{}}
\hline\hline
\multicolumn{1}{c}{name} & \multicolumn{1}{c}{material} & \multicolumn{1}{c}{thickness} & \multicolumn{1}{c}{density} & \multicolumn{1}{c}{$C_{11}^-$} & \multicolumn{1}{c}{$C_{11}^+$} & \multicolumn{1}{c}{$C_{44}^-$} & \multicolumn{1}{c}{$C_{44}^+$} \\
\multicolumn{1}{c}{} & \multicolumn{1}{c}{} & \multicolumn{1}{c}{(mm)} & \multicolumn{1}{c}{$\mathrm{(g/cm^3)}$} & \multicolumn{4}{c}{(Gpa)} \\ \hline
$\textbf{Si$<$100$>$}$ & Silicon & 0.28 & 2.33 & 165.6 & 203.9 & 50.9 & 79.5 \\
$\textbf{Si$<$111$>$}$ & \textemdash & 1.00 & \textemdash & \textemdash & \textemdash & \textemdash & \textemdash \\
$\textbf{Zry-a}$ & Zircaloy-4 & 1.33 & 6.50 & 137.6 & 165.5 & 32.0 & 48.9 \\
$\textbf{Zry-b}$ & \textemdash & 5.00 & \textemdash & \textemdash & \textemdash & \textemdash & \textemdash \\ \hline\hline
\end{tabular}
\end{adjustbox}
\end{table}
 
\subsection{Comparison techniques}
The Si samples, being single crystals, can be compared with the elastic constants taken from the literature \citep{Hall1967}. The correct orientation, i.e. the ${<}100{>}$ or ${<}111{>}$ direction normal to the sample surface, was obtained by rotating the reference tensor.

For the Zry samples, X-ray diffraction and neutron diffraction where carried for the "thin" ($\textbf{Zry-a}$) and "thick" ($\textbf{Zry-b}$) samples respectively. These measurements resulted in sets of pole figures which were inserted into MTEX \citep{Bachmann2010} to compute an Orientation Distribution Function, from which the texture coefficients were extracted. This coefficients were subsequently used under the framework developed by Lobos \citep{lobos2018homogenization}, together with the Zr single crystal elasticity \citep{Fisher1964}, to estimate the macroscopic elastic tensor.

All ultrasonic measurements where made without aligning the principal directions with the experimental directions. Therefore, all comparison tensors were subsequently rotated in the azimuthal direction to align with the orientation in the ultrasonic measurement.

\section{Results}\label{sec:4}
\subsection{Fitting agreement}\label{subsec:4:1}
\begin{figure}
    \centering
    \includegraphics[width=0.95\linewidth]{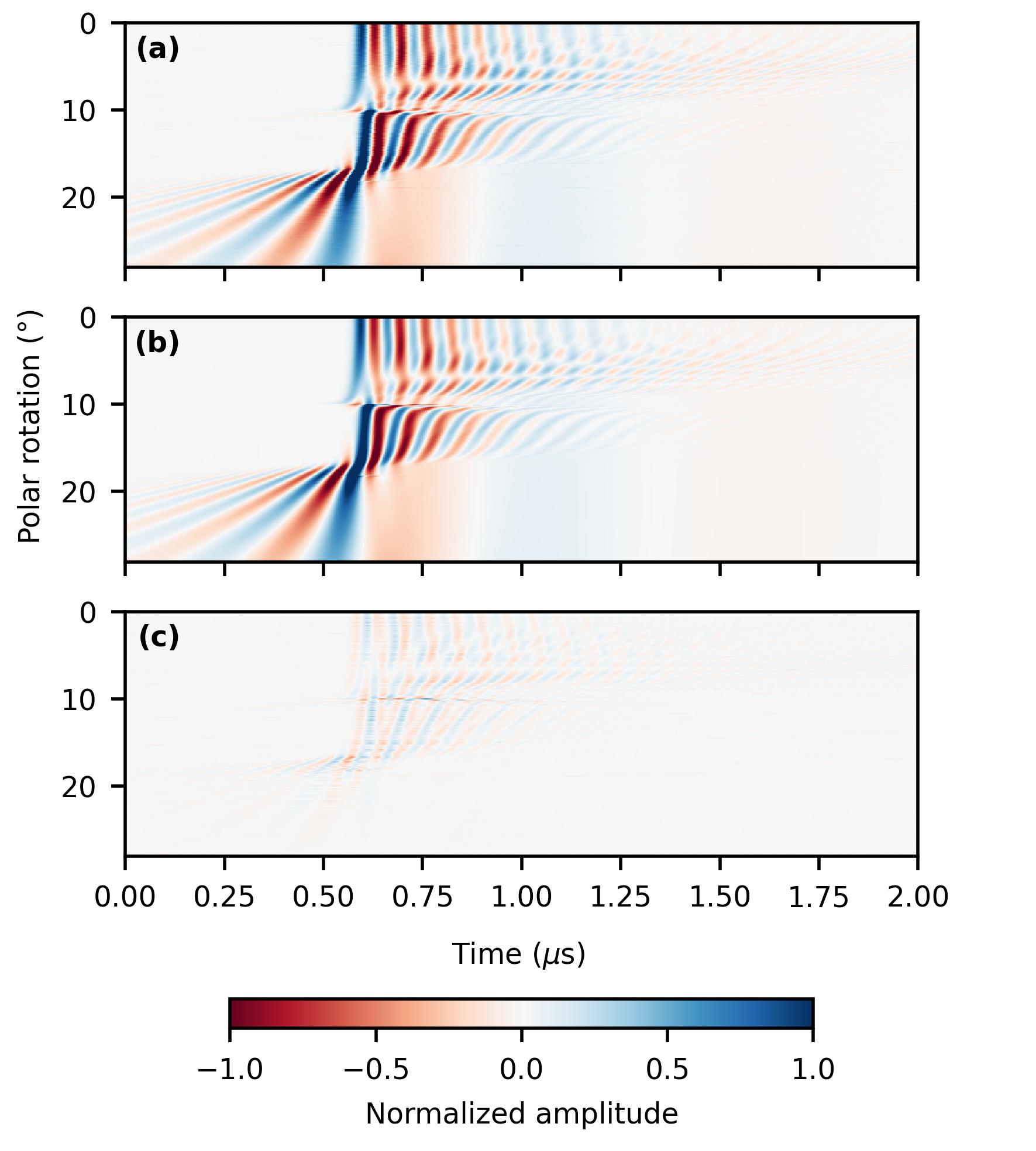}
    \caption{Comparison of experimental and fitted signals, for a single polar scan ( $\varphi=10^{\circ}$ ) in the Si sample. (a) experimental signal. (b) fitted signal. (c) subtraction of (a) and (b). (Color online).}
    \label{Figure5}
\end{figure}

\begin{figure}
    \centering
    \includegraphics[width=0.95\linewidth]{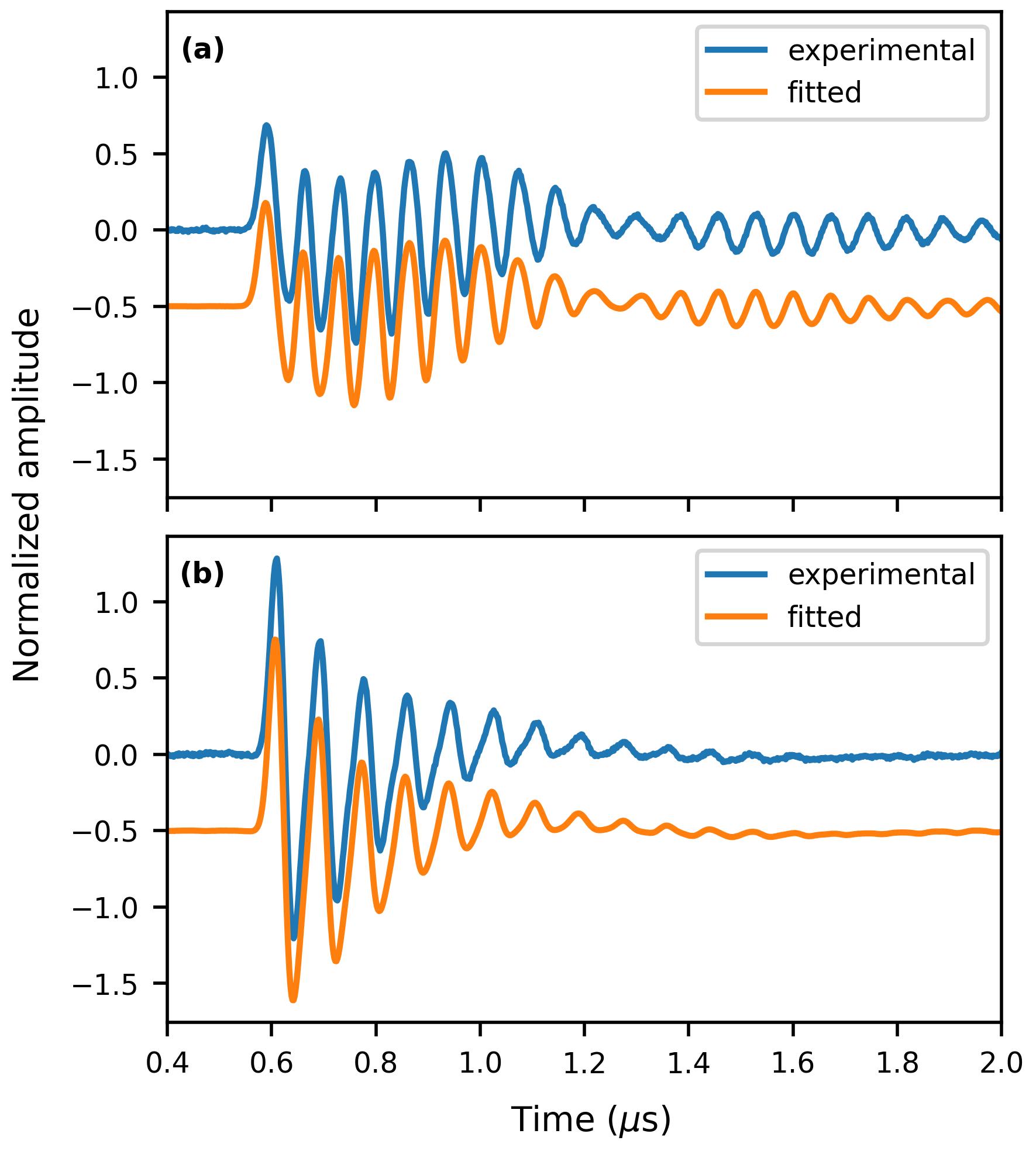}
    \caption{Comparison of experimental and fitted signals ( $\varphi=10^{\circ}$ ) for the Si sample. (a) $\theta=5.7^{\circ}$. (b) $\theta=11.3^{\circ}$. The fitted signals have an offset of -0.5 for better visualization. (Color online).}
    \label{Figure6}
\end{figure}

Agreement between the fitted and experimental data can be assessed by direct comparison of the datasets. This is shown for the \textbf{Si$<$100$>$} sample in Figs. \ref{Figure5} and \ref{Figure6}. Fig \ref{Figure5} shows the comparison for a polar scan taken at a single azimuthal plane. The low energy present in the subtraction (Fig. \ref{Figure5}.c) indicates good agreement. Fig. \ref{Figure6} displays superimposed signals at two different polar angles from the same azimuthal plane, again showing the similarity between experimental and fitted data. A comparison for a thicker sample is included in the supplementary material. Note that Fig. \ref{Figure6}.a shows an example of echo superposition which would void the bulk wave assumption often employed in the literature. The model described in subsection \ref{subsec:2:1} is well suited for this condition.

\subsection{Elastic constants}
The results are displayed in Table \ref{tab:elastic_grouped}, which include the stiffness matrix obtained by the ultrasonic method, the stiffness matrix obtained by the comparison method, the subtraction between the two, and the confidence intervals for the ultrasonic result. The latter were computed based on the covariance matrix following \citep{audoin1991estimation}. Table \ref{Table3} summarizes the results, showing the maximum absolute error and the total absolute error between the ultrasonic and comparison methods. The anisotropy factor is a measure of how anisotropic a material is, as defined in\citep{Lobos2016}. It is equivalent to the Zener factor\citep{zener1948elasticity} but applicable for any symmetry.

\begin{figure}
    \centering
    \includegraphics[width=0.9\linewidth]{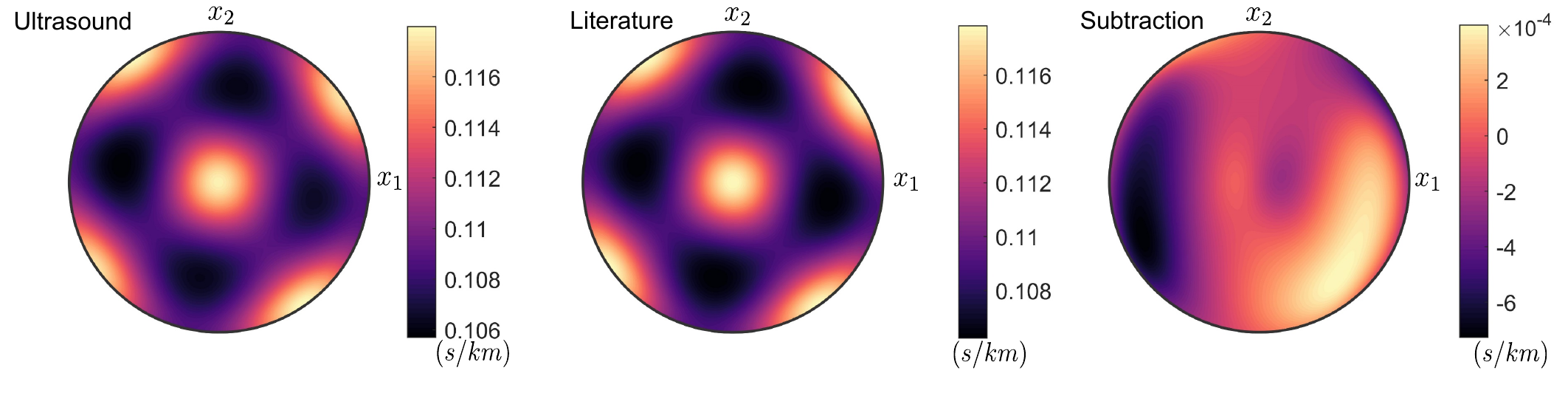}
    \caption{$\textbf{Si$<$100$>$}$}
    \label{fig:Fig7a}
\end{figure}

For a visual comparison, the slowness projections for the quasi-longitudinal mode where computed and plotted using MTEX. These are included in Fig. \ref{fig:Fig7a}, which include the projection for the ultrasonic result, the projection for the comparison method, and the subtraction between projections. Note that the sample principal directions were purposely misaligned with the experimental directions during measurement.


\newcolumntype{d}[1]{D{.}{.}{#1}}

\begin{table*}[!ht]
\centering
\caption{Stiffness matrices in Voigt notation, which include the ultrasonic result, the comparison method, the confidence interval of the ultrasonic method, and the subtraction between results. (a) \textbf{Si$<$100$>$}. (b) \textbf{Si$<$111$>$}. (c) \textbf{Zry-a}. (d) \textbf{Zry-b}.}
\label{Table2}

\begin{tabular}{@{}cccc@{}}
\multicolumn{4}{c}{(a) \textbf{Si$<$100$>$}}\\

{Ultrasound} &
{Literature} &
{Confidence interval} &
{Subtraction} \\[0.4em]

\resizebox{.24\textwidth}{!}{$
\left[
\begin{array}{d{3.1} d{3.1} d{3.1} d{3.1} d{3.1} d{3.1}}
191.6 & 38.8 & 64.2 & 0.3 & -1.5 & -9.2\\
&190.6 & 64.0 & 0.3 & -0.5 & 10.0\\
&&166.0 & 0.0 & 0.5 & 0.0\\
&&&79.5 & 0.1 & -0.1\\
&&\multicolumn{1}{c}{sim}&&79.5 & 0.0\\
&&&&&54.9
\end{array}
\right]
$}
&
\resizebox{.24\textwidth}{!}{$
\left[
\begin{array}{d{3.1} d{3.1} d{3.1} d{3.1} d{3.1} d{3.1}}
190.7 & 38.9 & 63.9 & 0.0 & 0.0 & -9.5\\
&190.7 & 63.9 & 0.0 & 0.0 & 9.5\\
&&165.6 & 0.0 & 0.0 & 0.0\\
&&&79.5 & 0.0 & 0.0\\
&&\multicolumn{1}{c}{sim}&&79.5 & 0.0\\
&&&&&54.4
\end{array}
\right]
$}
&
\resizebox{.24\textwidth}{!}{$
\left[
\begin{array}{d{3.1} d{3.1} d{3.1} d{3.1} d{3.1} d{3.1}}
1.1 & 0.9 & 0.8 & 1.2 & 1.1 & 0.4\\
&1.2 & 0.9 & 1.4 & 1.0 & 0.4\\
&&0.6 & 0.8 & 0.6 & 0.5\\
&&&0.4 & 0.3 & 0.5\\
&&\multicolumn{1}{c}{sim}&&0.3 & 0.4\\
&&&&&0.3
\end{array}
\right]
$}
&
\resizebox{.24\textwidth}{!}{$
\left[
\begin{array}{d{3.1} d{3.1} d{3.1} d{3.1} d{3.1} d{3.1}}
0.9 & -0.1 & 0.3 & 0.3 & -1.5 & 0.3\\
&-0.2 & 0.0 & 0.3 & -0.5 & 0.5\\
&&0.3 & 0.0 & 0.5 & 0.0\\
&&&0.0 & 0.1 & -0.1\\
&&\multicolumn{1}{c}{sim}&&0.0 & 0.0\\
&&&&&0.5
\end{array}
\right]
$}

\\
\\
\\
\multicolumn{4}{c}{(b) \textbf{Si$<$111$>$}}\\

{Ultrasound} &
{Literature} &
{Confidence interval} &
{Subtraction} \\[0.4em]

\resizebox{.24\textwidth}{!}{$
\left[
\begin{array}{d{3.1} d{3.1} d{3.1} d{3.1} d{3.1} d{3.1}}
193.6 & 54.2 & 45.2 & -13.1 & -0.9 & 0.2\\
&194.3 & 44.5 & 13.4 & 1.4 & 0.5\\
&&203.2 & 0.9 & 0.5 & 0.4\\
&&&60.0 & 0.2 & 1.2\\
&&\multicolumn{1}{c}{sym}&&60.3 & -13.4\\
&&&&&69.5
\end{array}
\right]
$}
&
\resizebox{.24\textwidth}{!}{$
\left[
\begin{array}{d{3.1} d{3.1} d{3.1} d{3.1} d{3.1} d{3.1}}
194.3 & 54.4 & 44.8 & -13.5 & -1.2 & 0.0\\
&194.3 & 44.8 & 13.5 & 1.2 & 0.0\\
&&203.8 & 0.0 & 0.0 & 0.0\\
&&&60.4 & 0.0 & 1.2\\
&&\multicolumn{1}{c}{sym}&&60.4 & -13.5\\
&&&&&70.0
\end{array}
\right]
$}
&
\resizebox{.24\textwidth}{!}{$
\left[
\begin{array}{d{3.1} d{3.1} d{3.1} d{3.1} d{3.1} d{3.1}}
1.1 & 0.7 & 0.6 & 0.7 & 0.8 & 0.5\\
&1.0 & 0.6 & 0.8 & 0.8 & 0.5\\
&&0.4 & 0.6 & 0.6 & 0.5\\
&&&0.2 & 0.2 & 0.3\\
&&\multicolumn{1}{c}{sym}&&0.2 & 0.2\\
&&&&&0.2
\end{array}
\right]
$}
&
\resizebox{.24\textwidth}{!}{$
\left[
\begin{array}{d{3.1} d{3.1} d{3.1} d{3.1} d{3.1} d{3.1}}
-0.7 & -0.2 & 0.4 & 0.3 & 0.3 & 0.2\\
&0.0 & -0.3 & 0.0 & 0.2 & 0.5\\
&&-0.6 & 0.9 & 0.5 & 0.4\\
&&&-0.4 & 0.2 & 0.0\\
&&\multicolumn{1}{c}{sym}&&-0.1 & 0.0\\
&&&&&-0.5
\end{array}
\right]
$}
\\
\\
\\
\multicolumn{4}{c}{(c) \textbf{Zry-a}}\\

{Ultrasound} &
{X-ray diffraction} &
{Confidence interval} &
{Subtraction} \\[0.4em]

\resizebox{.24\textwidth}{!}{$
\left[
\begin{array}{d{3.1} d{3.1} d{3.1} d{3.1} d{3.1} d{3.1}}
142.1 & 70.5 & 70.6 & 0.0 & -0.3 & 0.1\\
&142.5 & 71.1 & 0.0 & -0.2 & 0.4\\
&&151.1 & 0.0 & -0.1 & 0.2\\
&&&36.2 & 0.3 & -0.1\\
&&\multicolumn{1}{c}{sym}&&35.1 & 0.0\\
&&&&&35.7
\end{array}
\right]
$}
&
\resizebox{.24\textwidth}{!}{$
\left[
\begin{array}{d{3.1} d{3.1} d{3.1} d{3.1} d{3.1} d{3.1}}
144.1 & 71.6 & 69.7 & 0.0 & 0.0 & -0.1\\
&143.2 & 70.6 & 0.0 & 0.0 & -0.1\\
&&146.7 & 0.1 & 0.1 & 0.2\\
&&&35.4 & 0.2 & 0.0\\
&&\multicolumn{1}{c}{sym}&&34.7 & 0.0\\
&&&&&36.0
\end{array}
\right]
$}
&
\resizebox{.24\textwidth}{!}{$
\left[
\begin{array}{d{3.1} d{3.1} d{3.1} d{3.1} d{3.1} d{3.1}}
0.8 & 0.6 & 0.4 & 0.4 & 0.4 & 0.2\\
&0.4 & 0.2 & 0.4 & 0.4 & 0.2\\
&&0.2 & 0.4 & 0.3 & 0.2\\
&&&0.0 & 0.0 & 0.2\\
&&\multicolumn{1}{c}{sym}&&0.1 & 0.2\\
&&&&&0.3
\end{array}
\right]
$}
&
\resizebox{.24\textwidth}{!}{$
\left[
\begin{array}{d{3.1} d{3.1} d{3.1} d{3.1} d{3.1} d{3.1}}
-1.9 & -1.1 & 0.8 & 0.0 & -0.2 & 0.3\\
&-0.7 & 0.5 & 0.0 & -0.2 & 0.5\\
&&4.4 & -0.1 & -0.2 & 0.0\\
&&&0.7 & 0.1 & 0.0\\
&&\multicolumn{1}{c}{sym}&&0.5 & 0.0\\
&&&&&-0.3
\end{array}
\right]
$}

\\
\\
\\

\multicolumn{4}{c}{(d) \textbf{Zry-b}}\\

{Ultrasound} &
{Neutron diffraction} &
{Confidence interval} &
{Subtraction} \\[0.4em]

\resizebox{.24\textwidth}{!}{$
\left[
\begin{array}{d{3.1} d{3.1} d{3.1} d{3.1} d{3.1} d{3.1}}
143.9 & 71.3 & 71.3 & 0.4 & 0.1 & -0.2\\
&141.9 & 69.3 & 0.2 & 0.3 & 0.4\\
&&148.6 & 0.5 & 0.2 & -0.1\\
&&&35.0 & -0.4 & -0.1\\
&&\multicolumn{1}{c}{sym}&&38.1 & -0.1\\
&&&&&34.9
\end{array}
\right]
$}
&
\resizebox{.24\textwidth}{!}{$
\left[
\begin{array}{d{3.1} d{3.1} d{3.1} d{3.1} d{3.1} d{3.1}}
142.8 & 70.9 & 72.0 & 0.0 & -0.0 & -0.4\\
&142.2 & 69.7 & -0.1 & -0.0 & 0.3\\
&&147.7 & 0.2 & 0.0 & -0.3\\
&&&34.5 & -0.4 & -0.0\\
&&\multicolumn{1}{c}{sym}&&37.5 & 0.1\\
&&&&&34.8
\end{array}
\right]
$}
&
\resizebox{.24\textwidth}{!}{$
\left[
\begin{array}{d{3.1} d{3.1} d{3.1} d{3.1} d{3.1} d{3.1}}
0.4 & 0.4 & 0.2 & 0.4 & 0.2 & 0.3\\
&0.8 & 0.4 & 0.2 & 0.2 & 0.3\\
&&0.1 & 0.2 & 0.2 & 0.3\\
&&&0.1 & 0.1 & 0.2\\
&&\multicolumn{1}{c}{sym}&&0.1 & 0.1\\
&&&&&0.2
\end{array}
\right]
$}
&
\resizebox{.24\textwidth}{!}{$
\left[
\begin{array}{d{3.1} d{3.1} d{3.1} d{3.1} d{3.1} d{3.1}}
1.1 & 0.4 & -0.7 & 0.4 & 0.1 & 0.2\\
&-0.3 & -0.4 & 0.3 & 0.3 & 0.0\\
&&1.0 & 0.3 & 0.2 & 0.2\\
&&&0.5 & -0.0 & -0.1\\
&&\multicolumn{1}{c}{sym}&&0.6 & -0.1\\
&&&&&0.1
\end{array}
\right]
$}

\end{tabular}

\label{tab:elastic_grouped}
\end{table*}

\begin{figure}
    \centering
    \includegraphics[width=0.9\linewidth]{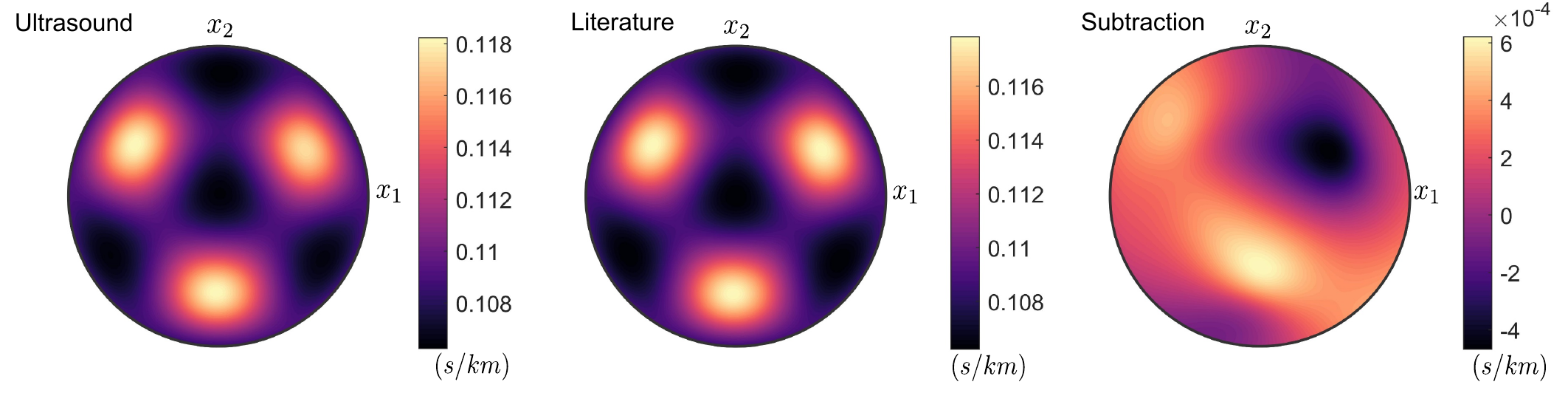}
    \caption{$\textbf{Si$<$111$>$}$}
    \label{fig:Fig7b}
\end{figure}
\begin{figure}
    \centering
    \includegraphics[width=0.9\linewidth]{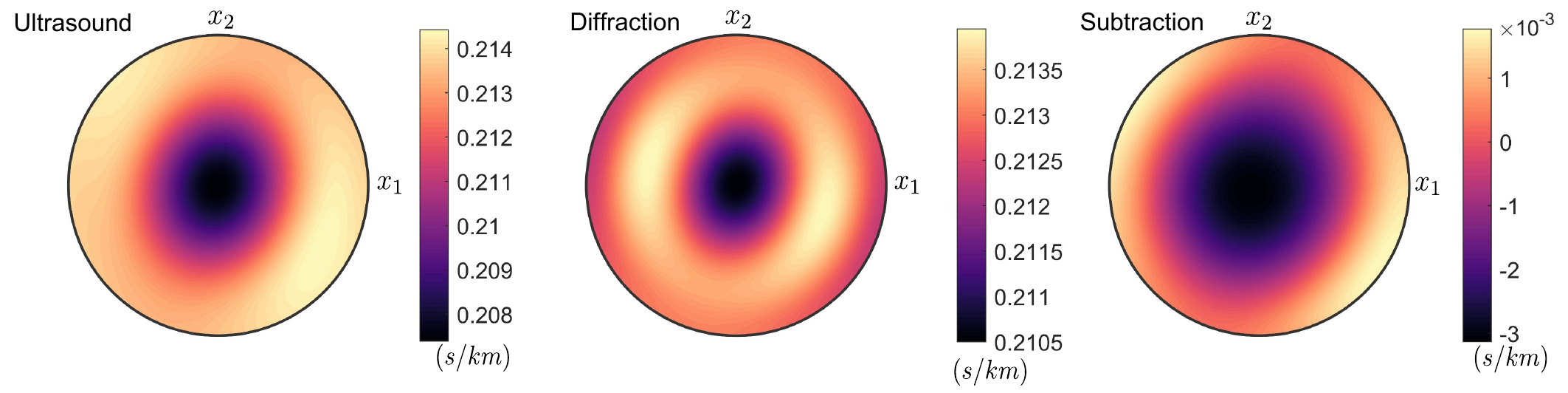}
    \caption{$\textbf{\textbf{Zry-a}}$}
    \label{Fig7b}
\end{figure}
\begin{figure}
    \centering
    \includegraphics[width=0.9\linewidth]{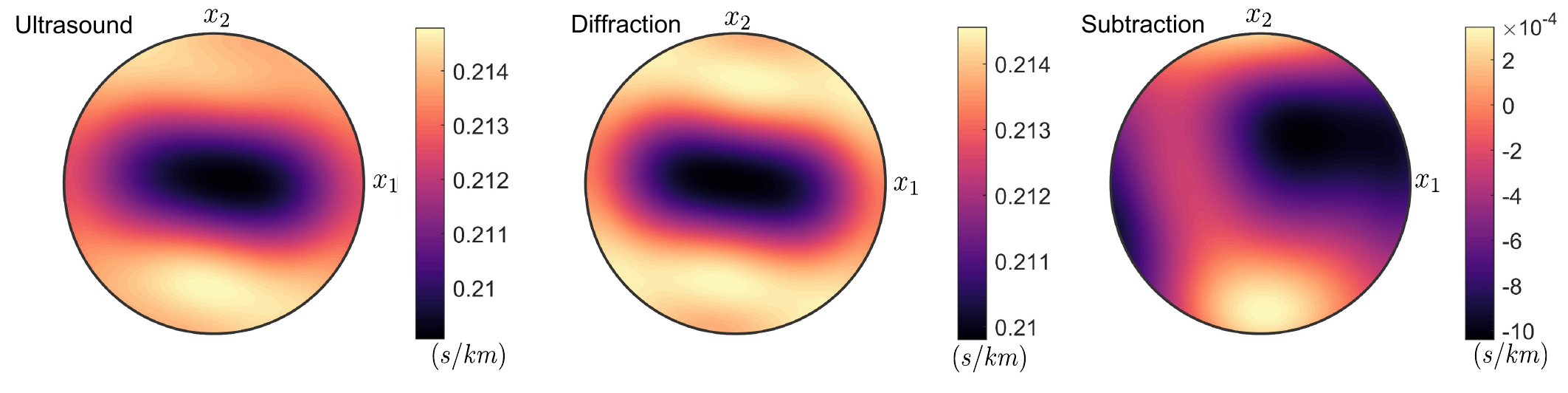}
    \caption{$\textbf{\textbf{Zry-b}}$}
    \label{Fig7d}
\end{figure}


\begin{table}[]
\caption{Additional Results.}
\label{Table3}
\begin{ruledtabular}
\begin{tabular}{@{}ccccc@{}}

\multicolumn{1}{l}{} & \multicolumn{1}{c}{$\textbf{Si$<$100$>$}$} & \multicolumn{1}{c}{$\textbf{Si$<$111$>$}$}& \multicolumn{1}{c}{$\textbf{Zry-a}$} & \multicolumn{1}{c}{$\textbf{Zry-b}$} \\ \hline
Max error (Gpa)& 1.5 & 0.9 & 4.4 & 1.1 \\
Total error (Gpa)& 6.4 & 6.8 & 12.5 & 7.4 \\
Anisotropy factor& 1.61 & 1.61 & 1.20 & 1.18 \\ 
\end{tabular}
\end{ruledtabular}

\end{table}

\section{Discussion}\label{sec:5}
Remarkable agreement was found for the Silicon samples, between the ultrasonic method and the reference values taken from the literature. The \textbf{Si$<$100$>$} wafer confirms the applicability of the method to thin specimens, as the model accounts for both fluid-solid interfaces, and the corresponding reflected, transmitted, and refracted waves. Moreover, the similarity between the measured and fitted signals, included in subsection \ref{subsec:4:1}, show that the experimental method, mainly due to the ad hoc transducers, sufficiently resembles plane wave propagation. 

The \textbf{Si$<$111$>$} wafer does not have in-plane mirror symmetry, which is a common assumption (and limitation) of the literature. Additionally, the purposely added azimuthal rotation forces the misalignment between the symmetry axis ($x_1',\ x_2'$) and the experimental axis ($x_1,\ x_2$). The triclinic assumption makes the method applicable under these conditions proving the technique is applicable for virtually any symmetry. An intrinsic limitation is discussed in the appendix for macroscopically hexagonal materials.

Moreover, the single crystals, are analogous to a polycrystalline material with a very sharp crystallographic texture. These results shows that even using the isotropic initial guess, the optimization algorithm is capable of converging to a different condition near the edge of the search space, defined by the proposed bounds. This attribute may depend on the single crystal anisotropy, and should be confirmed for systems with a higher intrinsic anisotropy. Table \ref{Table3} shows that the anisotropy factor is relatively low for the measured samples.

Good agreement was also found for the two Zry samples, between the ultrasonic method and the diffraction based methods. These results confirm that the method is applicable for samples of various thicknesses. When comparing these samples, a lower error was observed for the "thick" sample (\textbf{Zry-b}) than for the "thin" sample (\textbf{Zry-a}), as show in Table \ref{Table3}. This may correspond to the fact that neutron diffraction is a bulk transmission technique due to the high penetration of the neutron beam, whereas X-ray diffraction is a shallow reflection technique. Therefore, the ultrasonic method, being volumetric, is expected to better resemble the former. 

While a great advantage of the current implementation is that there is no need to know the macroscopic symmetry, nor to align symmetry directions, the initial guess and bounds are computed using the microscopic (single crystal) elasticity tensor. This means that the best usage of this method is to characterize known materials with unknown microstructure, rather than completely unknown materials. This can be particularly useful for processes like additive manufacturing where the resulting microstructure may change while changing fabrication parameters. When compared to the traditional approach, where the initial guess is taken from previously measured materials \citep{Kim2009}, and the bounds are defined by multiplying the initial guess by fixed factors \citep{Martens2019}, the method proposed here ensures that the search space encompass all possible realizations of the material under investigation, and the initial guess is a self consistent solution which does not rely on previous measurements.  

Efforts in the reduction in time were carried with the automation of the experiment and with the GPU implementation of the forward model. This resulted in the measurement and inversion of a single sample in under ten minutes. 

\section{Conclusion}\label{sec:6}
This work investigated the ultrasonic determination of elastic constants using an oblique-incidence immersion methodology. In contrast to resonance-based techniques, this approach requires only a specimen with two parallel faces, resulting in broad applicability and minimal sample preparation. Unlike shallow methods, like surface acoustic waves, the transmission configuration enables direct access to bulk elastic properties. This capability likely accounts for the closer agreement observed with neutron diffraction measurements compared to X-ray diffraction results.

The implementation of a forward model that explicitly accounts for fluid–solid interfaces enables robust fitting over a wide range of sample thicknesses, including cases where individual echoes cannot be resolved. This feature allows the use of lower excitation frequencies while preserving the medium homogeneity approximation, which is often violated at higher frequencies due to increased scattering. Furthermore, the adoption of a triclinic symmetry assumption permits the characterization of materials with arbitrary symmetry and relaxes the requirement for precise sample alignment.

The strong agreement between fitted and experimental signals confirms that, owing to the use of specifically designed transducers, the plane-wave approximation is adequate for the present configuration, obviating the need for computationally intensive finite-beam models. This simplification is particularly significant given the large number of forward evaluations required by the derivative-free optimization procedure. Computational cost was further reduced through a GPU-based implementation of the forward model, enabling full elastic inversions in under ten minutes.

Finally, the use of optimal zeroth-order bounds effectively constrains the search space, while the isotropic self-consistent solution provides a reliable initial estimate without depending on prior experimental results. The method does, however, require knowledge of the microscopic elasticity tensor. Consequently, it is best suited for the characterization of materials with known microscopic elastic properties but unknown microstructure, rather than for fully unknown materials.

\appendix

 \section{Uniqueness}
The Christoffel equation establishes a one to one correspondence between the  stiffness tensor, and the phase velocities and wave polarizations, i. e. for a given stiffness tensor only a set of phase velocities and wave polarizations exist, and vice versa. Nevertheless, if only the phase velocities are taken into account, and the polarizations are disregarded, the bijectivity disappears due to the phenomenon called anomalous partners. This meas that for two different stiffness tensors, the set of phase velocities computed with the Christoffel equation are the same \citep{Helbig2009}. Because the present work employs a waveform fitting approach, instead of velocity parametrization, the method naturally includes wave polarization in the computation of the transmission coefficients and therefore cannot concur in the anomalous partner error. Moreover, the use of zero-th order bounds limits the search space, which further reduces the risk for that error.
On the other hand, in oblique incidence immersion experiments, the shear horizontal mode can only be excited in non symmetry planes. This only represents a problem if, for the material under investigation, all planes are symmetry planes. This is the case for hexagonal materials with the symmetry axis being the plate normal ($x_3'$). The Navier equation for this material results in
\begin{equation}
\begin{gathered}
    C_{11}\frac{\partial^2 u_1}{\partial x_1^2}
    +C_{44}\frac{\partial^2 u_1}{\partial x_3^2}+
    (C_{13}+C_{44})\frac{\partial^2 u_3}{\partial x_1\partial x_3}=\rho \ddot{u}_1\\
        \frac{C_{11}-C_{12}}{2}\frac{\partial^2 u_2}{\partial x_1^2}
    +C_{44}\frac{\partial^2 u_2}{\partial x_3^2}=\rho \ddot{u}_2\\
        (C_{13}+C_{44})\frac{\partial^2 u_1}{\partial x_1\partial x_3}    
    +C_{44}\frac{\partial^2 u_3}{\partial x_1^2}
    +C_{33}\frac{\partial^2 u_3}{\partial x_3^2}
    =\rho \ddot{u}_3  
\end{gathered}
\end{equation}
It can be seen that, due to uncoupling, the Shear Horizontal mode cannot be excited, and because the $C_{12}$ constant depends only on horizontal movement ($u_2$), it cannot be inverted. This information could be retrieved, for example, by using a shear wave contact transducer \citep{Aristgui1997}. This issue also applies to perfectly isotropic materials, for which the incorrect determination of $C_{12}$ can lead to it being misclassified as hexagonal. Nevertheless, if the material is known to be isotropic that constant can be disregarded and replaced by $C_{13}$. For all other symmetries there exists a one to one relation between the elastic constants and the transmitted field.

\bibliography{references.bib}

\end{document}